\documentclass[aps,superscriptaddress,prl,twocolumn,nofootinbib]{revtex4}

\usepackage{graphicx}
\usepackage{epstopdf}
\usepackage{subfigure}
\usepackage{epsfig}
\usepackage{amsmath,amssymb,amsfonts}
\usepackage{color}
\usepackage{xcolor}
\usepackage{url}
\usepackage[utf8]{inputenc}
\usepackage[bookmarks,
                bookmarksopen = true,
                bookmarksnumbered = true,
                linktocpage,
                colorlinks = true,
                linkcolor = blue,
                urlcolor  = blue,
                citecolor = blue,
                anchorcolor = green,
                hyperindex = true,
                hyperfigures]
                {hyperref}

\begin{document}

\title{Constraining the equation of state with heavy quarks in the quasi-particle model of QCD matter}

\author{Feng-Lei Liu}
\email{flliu@mails.ccnu.edu.cn}
\affiliation{Institute of Particle Physics and Key Laboratory of Quark and Lepton Physics (MOE), Central China Normal University, Wuhan, Hubei 430079, China}

\author{Xiang-Yu Wu}
\email{xiangyuwu@mails.ccnu.edu.cn}
\affiliation{Institute of Particle Physics and Key Laboratory of Quark and Lepton Physics (MOE), Central China Normal University, Wuhan, Hubei 430079, China}

\author{Shanshan Cao}
\email{shanshan.cao@sdu.edu.cn}
\affiliation{Institute of Frontier and Interdisciplinary Science, Shandong University, Qingdao, Shandong 266237, China}

\author{Guang-You Qin}
\email{guangyou.qin@ccnu.edu.cn}
\affiliation{Institute of Particle Physics and Key Laboratory of Quark and Lepton Physics (MOE), Central China Normal University, Wuhan, Hubei 430079, China}

\author{Xin-Nian Wang}
\email{xnwang@lbl.gov}
\thanks{Current address: 3.}
\affiliation{Institute of Particle Physics and Key Laboratory of Quark and Lepton Physics (MOE), Central China Normal University, Wuhan, Hubei 430079, China}
\affiliation{Nuclear Science Division, Lawrence Berkeley National Laboratory, Berkeley, CA 94720, USA}

\date{\today}


\begin{abstract}

{In a quasi-particle model of QCD matter at finite temperature with thermal masses for quarks and gluons from hard thermal loops, the equation of state (EOS) can be described by an effective temperature dependence of the strong coupling $g(T)$. Assuming the same effective coupling between the exchanged gluon and thermal partons, the EOS can also be related to parton energy loss.}
Based on the quasi-particle linear Boltzmann transport (QLBT) model coupled to a (3+1)-dimensional viscous hydrodynamic model of the quark-gluon plasma (QGP) evolution and a hybrid fragmentation-coalescence model for heavy quark hadronization, we perform a Bayesian analysis of the experimental data on $D$ meson suppression $R_{\rm AA}$ and anisotropy $v_2$ at RHIC and the LHC. We achieve a simultaneous constraint on the QGP EOS and the heavy quark transport coefficient, both consistent with the lattice QCD results.

\end{abstract}

\maketitle


{\em Introduction.}
Since the discovery of the quark-gluon plasma (QGP)~\cite{Gyulassy:2004zy,Jacobs:2004qv} at the beginning of this century, the precise extraction of its properties has been one of the key goals of the relativistic heavy-ion collision programs at RHIC and the LHC~\cite{Busza:2018rrf}. It is now generally accepted that the QGP in heavy-ion collisions is a strongly coupled system that behaves like a nearly perfect fluid. This allows a hydrodynamic description of the QGP~\cite{Heinz:2013th,Jeon:2015dfa}, based on which one may obtain crucial properties of the system from experimental data on the low transverse momentum (soft) hadrons emitted from the QGP with the help of Bayesian interference. These include the shear and bulk viscosities of the QGP~\cite{Novak:2013bqa,Bernhard:2019bmu,JETSCAPE:2020shq,Nijs:2020ors}, and its equation of state (EoS)~\cite{Pratt:2015zsa}; the latter agrees well with the lattice QCD results.

Jets and heavy quarks, known as ``hard probes", provide an alternative way of revealing the QGP properties~\cite{Majumder:2010qh,Qin:2015srf,Dong:2019byy,Rapp:2018qla}. They are produced from the primordial hard collisions and observe the full evolution history of the QGP. The significant quenching of their spectra at high transverse momentum ($p_\mathrm{T}$) in nucleus-nucleus compared to proton-proton collisions~\cite{PHENIX:2006iih,PHENIX:2008saf} is deemed a smoking gun evidence of the quark and gluon degrees of freedom at high temperature. Tremendous efforts have been devoted to understanding the elastic~\cite{Qin:2007rn,Moore:2004tg,Gossiaux:2008jv,Riek:2010fk,He:2011qa,Cao:2012jt,Das:2015ana,Song:2015ykw,Scardina:2017ipo} and inelastic~\cite{Gyulassy:1993hr,Wang:1991xy,Zakharov:1997uu,Baier:1998yf,Wang:2001ifa,Arnold:2002ja,Gyulassy:2003mc,Kovner:2003zj,Majumder:2009ge,Cao:2013ita,Uphoff:2014hza,Nahrgang:2013saa,Ke:2018tsh,Liu:2020dlt} energy loss of jet partons through the QGP. In addition to the medium modification of jets, the jet-induced medium excitation is also found essential for understanding the jet observables, especially how the lost energy of partons is redistributed in the phase space~\cite{Cao:2020wlm, Cao:2022odi}. These lead to a more comprehensive picture of jet-medium interactions, and extend the related observables from single inclusive hadrons~\cite{Vitev:2002pf,Salgado:2003gb,Dainese:2004te,Wicks:2005gt,Armesto:2005iq,Bass:2008rv, Armesto:2009zi,Renk:2011gj,Horowitz:2011gd,Chen:2011vt,Cao:2017hhk,Xing:2019xae}, di-hadron~\cite{Majumder:2004pt,Zhang:2007ja,Renk:2008xq,Cao:2015cba}, to full jets~\cite{Aad:2014bxa, Khachatryan:2016jfl,Qin:2010mn,Young:2011qx,Dai:2012am,Wang:2013cia,Blaizot:2013hx,Mehtar-Tani:2014yea,Cao:2017qpx,Kang:2017frl,He:2018xjv,He:2022evt,JETSCAPE:2022jer}, jet structures~\cite{Ramos:2014mba,Lokhtin:2014vda,Chien:2015hda,Casalderrey-Solana:2016jvj,Tachibana:2017syd,KunnawalkamElayavalli:2017hxo,Park:2018acg,Luo:2018pto,Chang:2017gkt,Mehtar-Tani:2016aco,Milhano:2017nzm,Caucal:2019uvr,Chen:2020tbl} and jet-related correlations~\cite{Aad:2010bu,Chatrchyan:2012gt,Qin:2009bk,Chen:2016vem,Chen:2016cof,Chen:2017zte,Zhang:2018urd,Kang:2018wrs,Yang:2021qtl,Luo:2021voy}.

The strength of jet-medium interactions can be quantified by a set of transport coefficients, such as the jet quenching parameter $\hat{q}$~\cite{Baier:2002tc,Majumder:2008zg} that characterizes the transverse momentum broadening of jet partons inside the QGP. Over the past decade, considerable efforts have been dedicated to developing theoretical and statistical tools for extracting $\hat{q}$ from the jet quenching data, ranging from a traditional $\chi^2$-fit method that determines its values at RHIC and LHC separately~\cite{JET:2013cls}, to a Bayesian statistical method that provides a continuous function of $\hat{q}$ with respect to the medium temperature and the parton energy~\cite{JETSCAPE:2021ehl}, and a recent information field based global Bayesian approach that starts with no specific ansatz of its functional form~\cite{Xie:2022ght}. Similarly, the diffusion coefficient of heavy quarks $D_\mathrm{s}$ has also been successfully obtained using various approaches and shown in agreement with the lattice QCD results~\cite{Xu:2017obm, Sambataro:2023tlv, Beraudo:2021ont, Liu:2016ysz, Ke:2018tsh}.

{In the above studies,  EOS parameterized from lattice QCD results,  transport coefficients and initial conditions determined from the soft hadron spectra are used in the viscous hydrodynamics for the QGP evolution. Jet transport coefficients are then constrained separately by jet quenching data with the QGP evolution given by the calibrated hydrodynamic models.} A closer investigation on the medium properties is performed in Ref.~\cite{Feal:2019xfl} where the density of scattering centers inside the QGP has been extracted from the jet quenching data and shown in qualitative consistency with the lattice EoS. In this study, we present {a direct Bayesian extraction of the QGP EOS using heavy flavor observables within the quasi-particle linear Boltzmann transport (QLBT) model \cite{Liu:2021dpm} which couples to the (3+1)-D CLVisc hydrodynamic model~\cite{Pang:2018zzo, Wu:2021fjf} of the QGP evoluton and a hybrid fragmentation-coalescence approach for heavy flavor hadronization~\cite{Cao:2019iqs}. In QLBT model, the EOS of the QCD matter can be described by an effective temperature dependence of the strong coupling $g(T)$ or the thermal mass of quasi-particles (quarks and gluons). Assuming the same effective coupling $g(T)$ between the exchanged gluon and thermal partons, the EOS can also be related to parton energy loss and therefore can be constrained by the jet quenching observables.} We can achieve a simultaneous constraint on QGP EOS and the heavy quark diffusion coefficient, both consistent with the lattice QCD results. {The shear viscosity from this constrained quasi-particle model is also consistent with the values used in CLVisc and from other Bayesian inferences.}

{\em Heavy quark interaction in QGP.}
{In the QLBT model~\cite{Liu:2021dpm},  QGP medium is viewed as a gas of massive quarks and gluons (quasi particles)~\cite{Das:2015ana,Song:2015ykw,Plumari:2011mk},} and the elastic and inelastic scatterings between heavy quarks and the medium are described by the linear Boltzmann transport model~\cite{Wang:2013cia,Cao:2016gvr}. The interaction among medium partons is encoded in their thermal masses \cite{Plumari:2011mk},
\begin{align}
\label{eq:thermalMass}
 & m_q^2(T) =\frac{N_c^2-1}{8N_c}g^2(T) T^2,\\
 \label{eq:thermalMass2}
 & m_g^2(T) =\frac{1}{6} \left(N_c+\frac{1}{2} N_f \right)  g^2(T) T^2,
\end{align}
for quarks and gluons respectively, where $N_c=3$ is the number of colors and $N_f=3$ is the number of flavors. {Since the interactions between the medium partons are at the thermal scale, we assume the coupling strength runs with the medium temperature as:}
\begin{align}
\label{eq:gT}
g^2(T)=\frac{48 \pi^2}{(11 N_c-2 N_f) \ln \left[\frac{\left(a T/T_\mathrm{c}+b\right)^2 }{1+c e^{-d (T/T_\mathrm{c})^2 }}\right]},
\end{align}
in which $T_\mathrm{c}$ denotes the transition temperature between the QGP and the hadronic matter, $a$, $b$, $c$ and $d$ are four parameters to be determined. Although the above equations are inspired by perturbative QCD (pQCD) calculations, it has been shown~\cite{Liu:2021dpm} that they are able to capture key features of the QCD EOS, which is non-perturbative in nature. However, unlike Ref.~\cite{Liu:2021dpm} where $g^2(T)$ is directly fitted from the lattice EOS, the inverse question is addressed in the present study -- whether one can constrain the QCD EOS with the heavy flavor observables in heavy-ion collisions.

Given thermal masses of these quasi-particles, interactions between them and heavy quarks (denoted by ``$a$") can be simulated via the Boltzmann equation,
\begin{align}
p_a&\cdot \partial f_a= \sum_{b c d }\int \prod_{i=b,c,d}\frac{d^3p_i}{2E_i(2\pi)^3} (f_cf_d-f_af_b)|{\cal M}_{ab\rightarrow cd}|^2\nonumber\\
 \times &\frac{\gamma_b}{2}
S_2(\hat s,\hat t,\hat u)(2\pi)^4\delta^4(p_a\!+\!p_b\!-\!p_c\!-\!p_d)+ {\rm inelastic},
\label{eq:Boltzmann}
\end{align}
where $p_{i}\, (i = a, b, c, d)$ is the four-momentum of parton $i$ ($b$ denotes the thermal parton with a spin-color degeneracy factor $\gamma_b$, $c$ and $d$ are the final states of $a$ and $b$ respectively), and $f_{i}$ is the phase-space distribution which is taken as $f_{i} = 1/(e^{p_{i}\cdot u / T} \pm 1)$ for thermal partons with the local temperature $T$ and flow velocity $u$ provided by the hydrodynamic simulation of the QGP evolution. The summation over $b$, $c$ and $d$ takes into account all possible elastic scattering processes whose scattering amplitudes $|{\cal M}_{ab \rightarrow cd}|^2$ are calculated at the leading order of pQCD~\cite{Combridge:1978kx}. A factor $S_2(\hat s,\hat t,\hat u)=\theta(\hat s\geq 2 \mu_\mathrm{D}^2)\theta(\hat t \leq-\mu_\mathrm{D}^2)\theta(\hat u \leq-\mu_\mathrm{D}^2)$ is imposed to regulate possible collinear divergence of $|{\cal M}_{ab \rightarrow cd}|^2$, with the Debye mass $\mu_\mathrm{D}^2(T) = 2m_g^2(T)$. As discussed in Ref.~\cite{Liu:2021dpm}, the thermal masses of medium partons enter in three locations of Eq.~(\ref{eq:Boltzmann}): the thermal distribution, the Mandelstam variables ($\hat s$, $\hat t$, $\hat u$), and the kinematic constraint ($S_2\times\delta^4$-function). The inelastic scattering process is related to the medium-induced gluon emission from heavy quarks, where the gluon spectrum is taken from the higher-twist energy loss calculation~\cite{Wang:2001ifa, Zhang:2003wk}. There are two types of coupling vertices in the heavy-quark-QGP interactions. {In order to simplify the Monte-Carlo simulation,} we assume the coupling between the exchanged gluon and thermal partons as the same as that between thermal partons in Eq.~(\ref{eq:gT}) that leads to the quasi-particle masses in Eqs.~(\ref{eq:thermalMass}) and (\ref{eq:thermalMass2}); {For vertices directly connecting with heavy quarks, we assume the interactions rely on the heavy quark energy scale and therefore parameterize the corresponding coupling strength as:}
\begin{align}
\label{eq:paraAlphas}
g^2(E)&=\frac{48 \pi^2}{(11N_c - 2N_f) \ln\left[\left(A  E/ T_\mathrm{c} + B\right)^2\right]},
\end{align}
with $A$ and $B$ as two additional parameters. Therefore, we have a total of six parameters in QLBT. {The energy of heavy quark here is evaluated in the local rest frame of the QGP medium. Note that Eq.~(3) can be reduced to Eq.~(5) if the energy scale of thermal partons is significantly larger than $T_\mathrm{c}$.}

We use the (3+1)-dimensional CLVisc hydrodynamic model~\cite {Pang:2018zzo,Wu:2018cpc} initialized with the AMPT model~\cite{Lin:2004en} to simulate the spacetime evolution of the QGP fireballs in heavy-ion collisions, from which quasi-particle distributions are extracted using the local temperature and flow velocity. The starting time of the hydrodynamic expansion ($\tau_0=0.6$~fm/$c$) and the specific shear viscosity $\eta/s=0.08$ have been adjusted to provide reasonable descriptions of the soft hadron spectra including the anisotropic flows. Heavy quarks are initialized using the Monte-Carlo Glauber model for their initial position and the LO pQCD calculations of the pair production and flavor excitation processes for their initial momentum distributions. Their subsequent interactions with the QGP are simulated using Eq.~(\ref{eq:Boltzmann}). On the chemical freeze-out hypersurface ($T_\mathrm{c}$), they are converted to hadrons using a hybrid coalescence-fragmentation hadronization model~\cite{Cao:2019iqs} that has been constrained by the heavy flavor hadron chemistry measurements. Interactions before ($\tau < \tau_0$) and after ($T<T_\mathrm{c}$) the QGP stage are neglected in this work.

\begin{figure*}[tbp]
\includegraphics[width=0.32\linewidth]{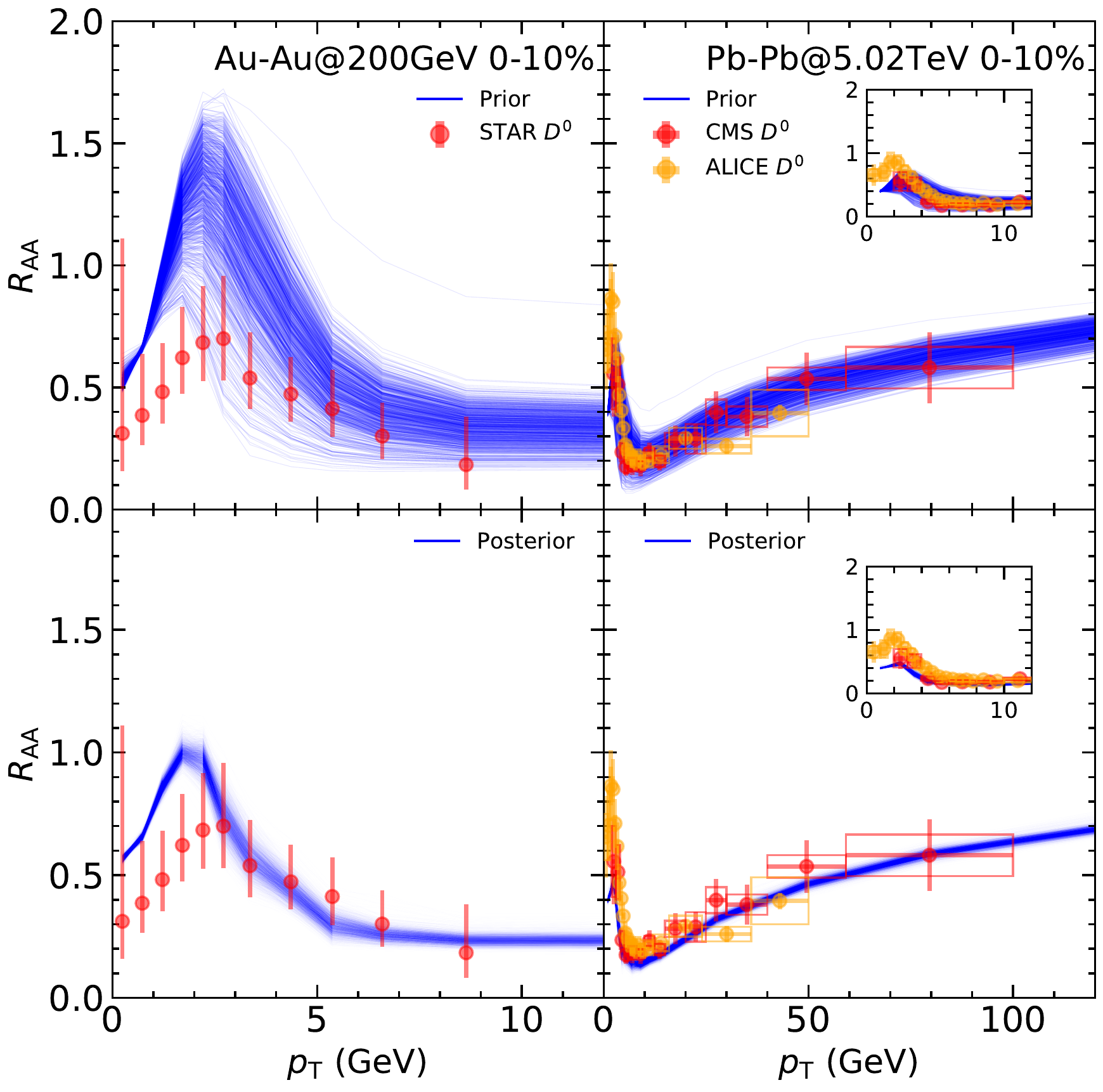}
\hspace{10pt}
\includegraphics[width=0.60\linewidth]{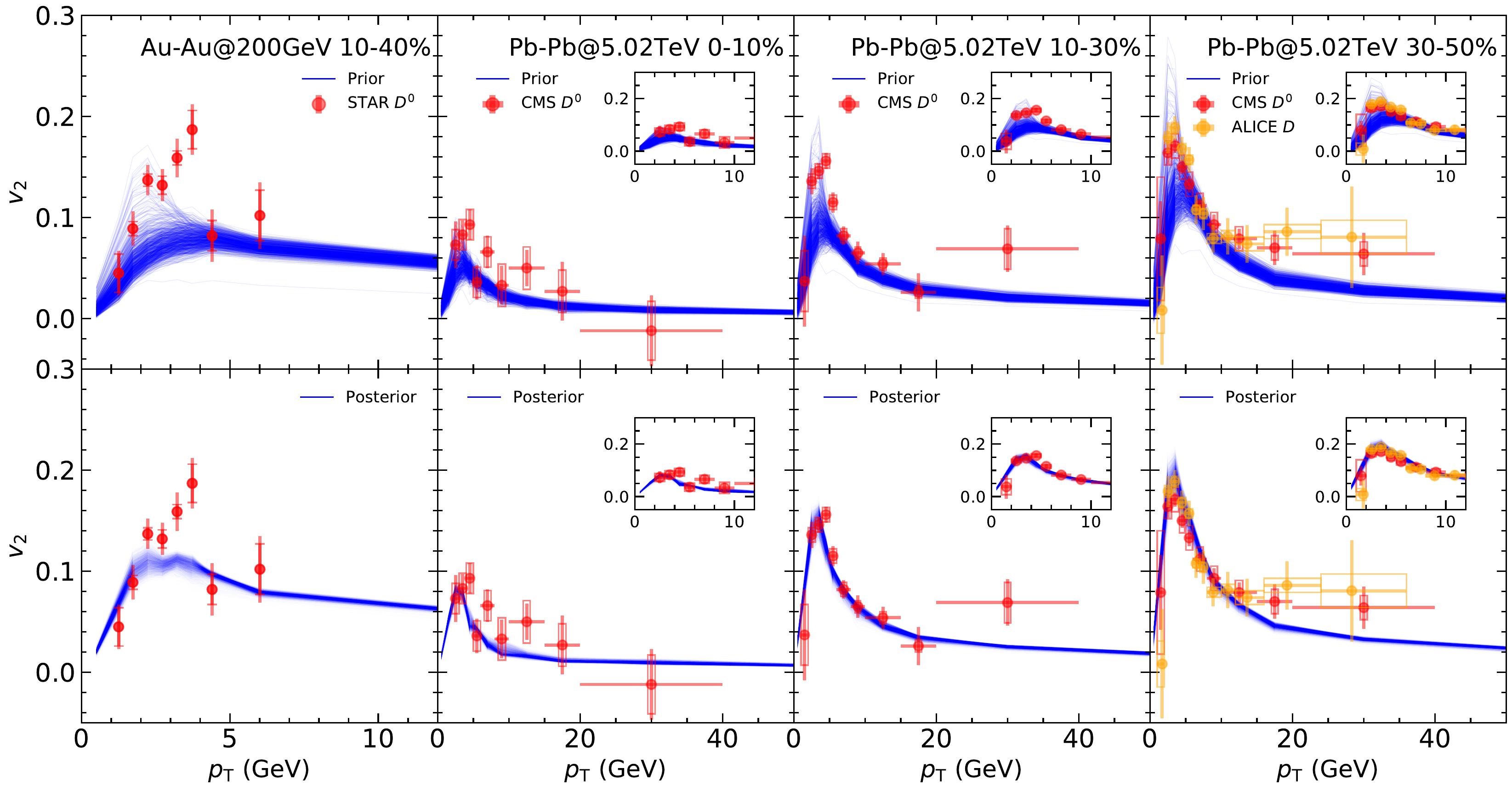}
\caption{(Color online) Calibration of the QLBT calculation (using $T_\mathrm{c}=150$~MeV) against the $D$ meson $R_\mathrm{AA}$ and $v_2$ data at RHIC~\cite{Adam:2018inb, Adamczyk:2017xur} and LHC~\cite{Sirunyan:2017xss, Sirunyan:2017plt}, upper panel for calculations with the prior distributions of parameters and lower panel with the posterior distributions after the calibration. {The inserted figures are $R_{\rm AA}$ and $v_2$ at the LHC in the same $p_{\rm T}$ range as RHIC results (i.e., $p_{\rm T} = 0\sim 12$~GeV).} {The ALICE data~\cite{ALICE:2021rxa} are also presented for comparison, though they have not been included in our model calibration.}}
\label{fig:RAA-v2}
\end{figure*}

{\em Bayesian extraction of the model parameters.}
To simultaneously constrain the six model parameters with the heavy flavor observables, we employ the Bayesian statistical analysis method, in which the posterior distribution of parameters ($\boldsymbol{\theta}=\{a,b,c,d,A,B\}$) given the experimental observation ($\mathrm{data}=\{y_i^\mathrm{exp}\}$) is proportional to the product of the likelihood of a given set of $\boldsymbol{\theta}$ and its distribution:
\begin{equation}
\label{eq:Bayesian}
P(\boldsymbol{\theta} | \mathrm{data}) \propto P(\mathrm{data} | \boldsymbol{\theta}) P(\boldsymbol{\theta}).
\end{equation}
The likelihood function can be estimated with a Gaussian form that quantifies the similarity between our model output ($\{y_i\}$) with a given  set of $\boldsymbol{\theta}$ and the data:
\begin{align}
\label{eq:likelihood}
P({\rm data} | \boldsymbol{\theta}) = \prod_i \frac{1}{\sqrt{2 \pi} \sigma_i} e^{- \frac{\left[y_i(\boldsymbol{\theta}) - y_i^\mathrm{exp}\right]^2}{2 \sigma_i^2}},
\end{align}
where the index $i$ runs over all data points included in our analysis, and $\sigma_i$ is the standard deviation at each data point that combines the experimental error and the interpolation error from our Gaussian process emulator (GP)~\cite{GaussianEmulator} which is trained with our QLBT calculations using over {800} sets of $\boldsymbol{\theta}$ and then applied as a fast surrogate of QLBT when we scan across the parameter space to constrain its posterior distribution.

With the above setup, the prior distribution of $\boldsymbol{\theta}$ is assumed uniform in the regions summarized in Tab.~\ref{tab:prior}. Note that in principle, the transition temperature $T_\mathrm{c}$ could also be a model parameter to be constrained from data. {However, in order to compare our extracted EoS to two different lattice QCD calculations -- Wuppertal–Budapest (WB)~\cite{Borsanyi:2013bia} and HotQCD (HQ)~\cite{Bazavov:2014pvz}, here we use two separate values of $T_\mathrm{c}$ from these two work -- 150~MeV for WB and 154~MeV for HQ.} {The ranges in Tab.~I are chosen such that they can produce a sufficiently wide prior range of the EoS with various shapes, a reasonable coverage of the experimental data from our model calculation, and a well constrained posterior distribution of the model parameters via the Bayesian calibration, as will be shown later.}

\begin{table}[htb]
\centering
\begin{tabular}{c|c|c}
 \hline
 \;\;Parameters\;\; & \begin{tabular}{@{}c@{}}Prior Range \\\;\; ($T_\mathrm{c}=150$~MeV)\;\;\end{tabular} &  \begin{tabular}{@{}c@{}}Prior Range \\\;\; ($T_\mathrm{c}=154$~MeV)\;\;\end{tabular}  \\
 \hline
   $a$ & [0.18, 4.5]      &  [0.26, 15.0] \\
   $b$ & [0.5, 1.2]      &  [0.1, 0.8]  \\
   $c$ & [-2.0, 5.0]    &  [-4.0, 20.0] \\
   $d$ & [0.35, 0.7]    &  [0.25, 0.65] \\
   $A$ & [0.05, 0.16]  &  [0.05, 0.18] \\
   $B$ & [0.6, 3.0]      &  [0.6, 3.2]\\
 \hline
\end{tabular}
\caption{The ranges of model parameters used in the prior distributions, for two different values of $T_\mathrm{c}$.}
\label{tab:prior}
\end{table}

\begin{figure}[tbp]
\includegraphics[width=0.93\linewidth]{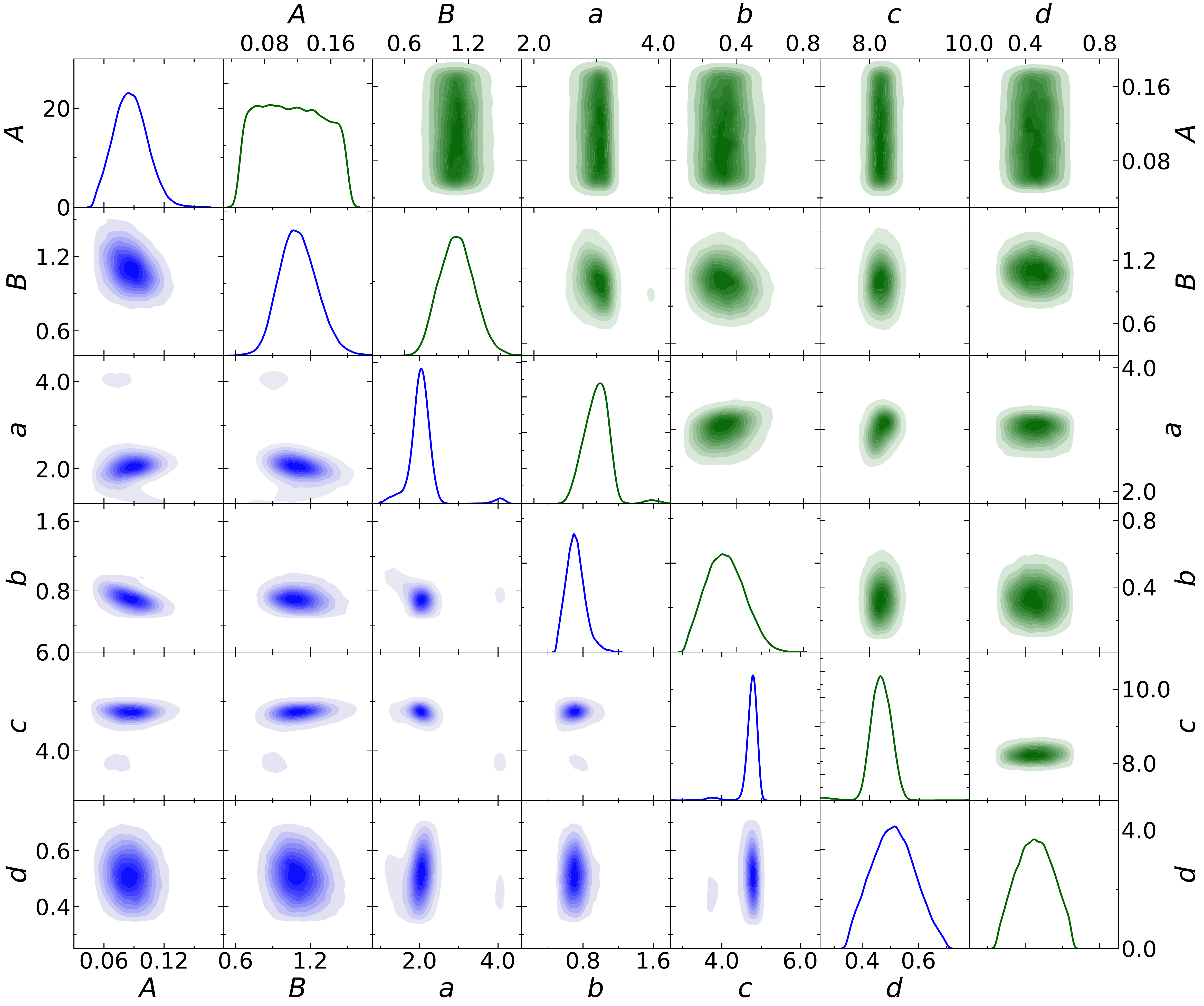}
\caption{(Color online) Posterior distributions of the model parameters, together with their correlations, upper triangle (green) for using $T_\mathrm{c}=154$~MeV, lower triangle (blue) for $T_\mathrm{c}=150$~MeV. }
\label{fig:posterior}
\end{figure}

\begin{figure*}[tbp]
\includegraphics[width=0.96\linewidth]{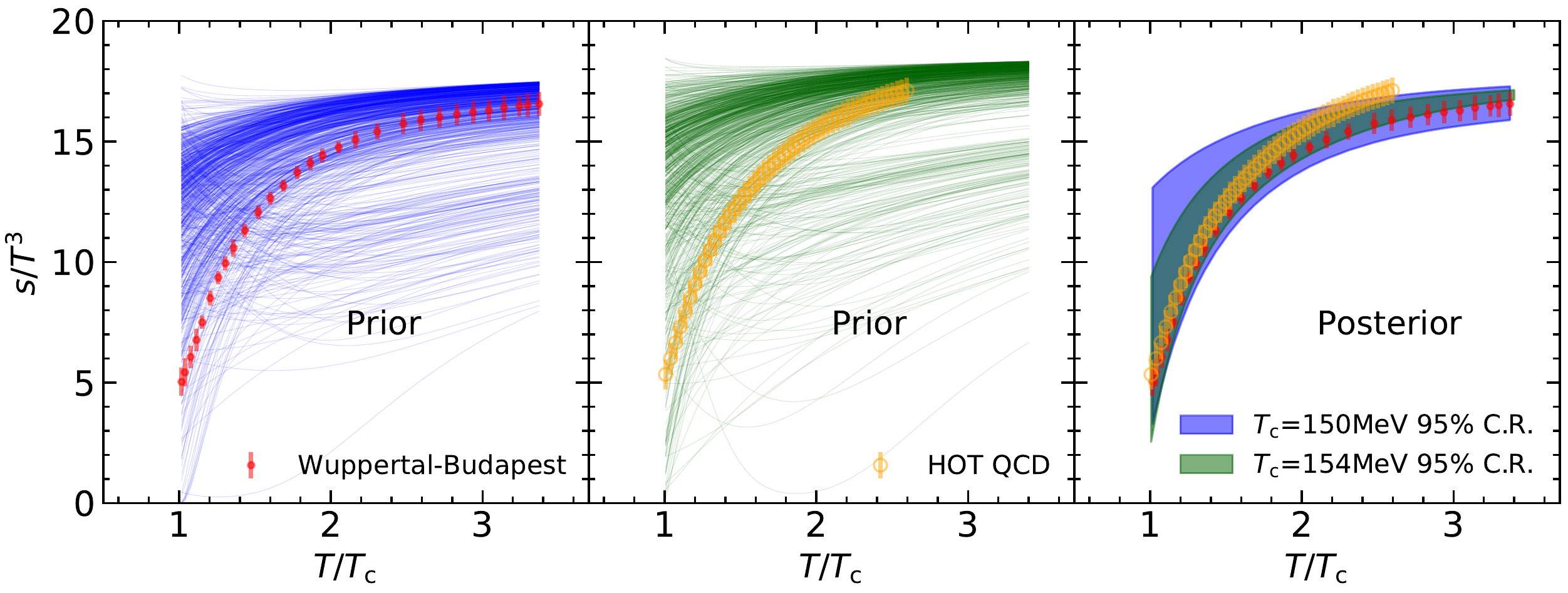}
\caption{(Color online) Entropy density as a function of temperature, left for its prior range mapped from Tab.~\ref{tab:prior} and right for its posterior range constrained by the heavy flavor observables, compared between results obtained with two different values of $T_\mathrm{c}$ and their corresponding lattice data~\cite{Borsanyi:2013bia,Bazavov:2014pvz}.}
\label{fig:EoS}
\end{figure*}

Starting with these prior distributions of $\boldsymbol{\theta}$, we calculate the nuclear modification factor $R_\mathrm{AA}$ and the elliptic flow coefficient $v_2$ of $D$ mesons using the QLBT model in  Au-Au collisions at 200~AGeV and Pb-Pb collisions at 5.02~ATeV with different centralities. The GP is trained with {821} parameter sets for the $T_\mathrm{c}=150$~MeV setup and {815} sets for $T_\mathrm{c}=154$~MeV. These design points are sampled according to the Latin-Hypercube algorithm~\cite{MORRIS1995381} and iterated for multiple rounds of trials to ensure both efficiency and stability in obtaining our results. With the GP, we employ the maximum a posterior (MAP) method in the PyMC library~\cite{pymc_bib} to perform the Markov Chain Monte Carlo (MCMC) estimation of the parameters with Metropolis-Hastings random walk in the parameter space according to the Bayesian statistics. We first implement {320,000} steps of random walks to approach the equilibrium of the posterior distribution, and then run another {320,000} steps from which we take {32,000} sets of parameters, each drawn from every 10 steps in order to exclude correlations from adjacent steps. These {32,000} sets represent the posterior distribution of $\boldsymbol{\theta}$ and are used to evaluate the QGP EoS in the end. {We use the integrated autocorrelation time method in emcee~\cite{Foreman-Mackey:2012any} to quantify the effects of sampling error on our results.
The integrated autocorrelation time $\tau_f$ is summarized in Tab.~\ref{tab:tau}.
We can see that the chains contain 32,000 sets of parameters, longer than 100 $\tau_f$ for all parameters.
This means that the chains are sufficiently converged.
}

\begin{table}[htb]
\centering
\begin{tabular}{c|c|c}
 \hline
 \;\;Parameters\;\; & \begin{tabular}{@{}c@{}}$\tau_f$ \\\;\; ($T_\mathrm{c}=150$~MeV)\;\;\end{tabular} &  \begin{tabular}{@{}c@{}}$\tau_f$ \\\;\; ($T_\mathrm{c}=154$~MeV)\;\;\end{tabular}  \\
 \hline
   $a$ & 178.242      &  33.871 \\
   $b$ & 16.175      &  2.187  \\
   $c$ & 310.174    &  40.206 \\
   $d$ &   1.329  &  0.967 \\
   $A$ & 3.764  &  1.069 \\
   $B$ & 14.365      &  1.119\\
 \hline
\end{tabular}
\caption{The integrated autocorrelation time $\tau_f$ of model parameters in the posterior distributions, for two different values of $T_\mathrm{c}$.}
\label{tab:tau}
\end{table}

Shown in Fig.~\ref{fig:RAA-v2} is the model calibration against the $D$ meson data at RHIC and LHC, left for $R_\mathrm{AA}$ and right for $v_2$. The upper panels are from our model calculations using the prior distributions of the parameters while the lower panels using their posterior distributions.  {One observes that after the Bayesian calibration, our QLBT model is able to provide a reasonable description of the $D$ meson observables in heavy-ion collisions. The remaining discrepancies between model and data may result from the limits existed in our ansatz of the coupling strength based on its perturbative form, and the energy loss and hadronization models.} Here we present results with $T_\mathrm{c}=150$~MeV. Similar results can be obtained  for $T_\mathrm{c}=154$~MeV. {We expect that heavy flavor observables at high $p_\mathrm{T}$ are roughly determined by the heavy quark energy loss, which is related to the overall magnitude of the couplings $g(T)$ and $g(E)$. However, at low to intermediate $p_\mathrm{T}$, these observables also depend on the medium flow, whose evolution profile is related to both the magnitude and the shape of the EoS. Therefore, comparing to high $p_\mathrm{T}$ region, heavy flavor observables at low to intermediate $p_\mathrm{T}$ region are more sensitive to the EoS.}

The extracted model parameters are presented in Fig.~\ref{fig:posterior}, in which the diagonal panels show their posterior distributions and off-diagonal panels show their correlations. Reasonable constraints on these model parameters have been obtained, and certain sensitivity of the extracted parameters on the value of $T_\mathrm{c}$ can also be observed in the figure (the upper triangle for $T_\mathrm{c}=154$~MeV and the lower triangle for $T_\mathrm{c}=150$~MeV). This will affect our constraint on the QGP EoS later.

{\em QGP EoS and transport coefficients.}
Using the six parameters constrained by data on the $D$ meson $R_\mathrm{AA}$ and $v_2$, we can evaluate the EoS of the quark-gluon plasma and the transport coefficients of heavy quarks and shear viscosity of the quasi-particle system.

With the temperature-dependent thermal masses of quarks and gluons that are determined by parameters $(a,b,c,d)$ via $g^2(T)$, one may calculate the pressure of the relativistic gas system as
\begin{equation}
  \label{eq:pressure}
  P(T) =\sum_{i} \gamma_i \int \frac{d^3 p}{(2 \pi)^3}
  \frac{p^2}{3 E_i(p, T)} f_i(p, T)-B(T),
\end{equation}
where we sum over all parton species with $E_i(p, T) = \sqrt{p^2 + m_i^2(T)}$ being their kinetic energy, $f_i(p, T)$ being the Bose/Fermi distribution for gluon/quark and $\gamma_i$ being their spin-color degeneracy factors. Similarly, the energy density is obtained as
\begin{equation}
  \label{eq:energyDensity}
  \epsilon(T) =\sum_{i} \gamma_i \int \frac{d^3 p}{(2\pi)^3}E_i(p, T) f_i(p, T)+B(T).
\end{equation}
Although a temperature-dependent bag term $B(T)$ is introduced in the above two equations to generate a phase transition between the QGP and the hadronic gas, it has no contribution to the entropy density: $s(T) = [\epsilon(T) + P(T)]/T$.
{Since this bag term cannot be determined from the coupling strength $g(T)$, it has not been included in the Boltzmann equation in this work.}

Shown in Fig.~\ref{fig:EoS} is the $T^3$-rescaled entropy density as a function of the temperature in the QGP phase. The left and middle panels show the variations of EOS from prior distributions of the parameters $(a,b,c,d)$ in the range given by Tab.~\ref{tab:prior}, while the right panel is obtained from their posterior distributions. One observes that the heavy flavor observables do provide constraints on the QGP EoS. Its 95\% Credible Region (C.R.) in the right panel collapses into two bands depending on the value of $T_\mathrm{c}$ we use. For a larger $T_\mathrm{c}$ that terminates the heavy-quark-QGP interaction earlier, a higher entropy density of the medium is required in order to produce the same amount of nuclear modification on heavy flavor hadrons observed by experiments. The EoS we extract with $T_\mathrm{c}=150$~MeV agrees well with the WB lattice data that shares the same $T_\mathrm{c}$, though some deviation can be observed from HQ data. We note that deviation exists between WB and HQ results on $s(T)/T^3$ as a function of either $T$ or $T/T_\mathrm{c}$.

\begin{figure}[tbp]
\includegraphics[width=0.99\linewidth]{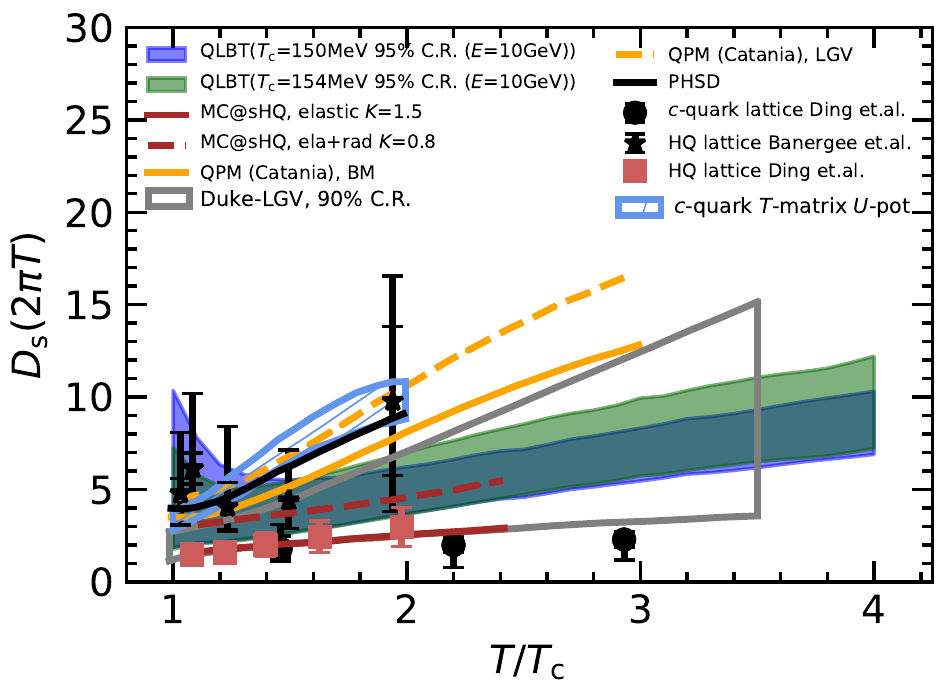}
\caption{(Color online) The spatial diffusion coefficient of charm quarks $D_\mathrm{s}$ as a function of the medium temperature, compared between extractions using two different values of $T_\mathrm{c}$ and results from other model calculations.}
\label{fig:Ds}
\end{figure}

One can also extract the interaction strength between heavy quarks and the QGP, which can be quantified by the transverse transport coefficient $\hat{q}$ of heavy quarks. It is defined as the transverse momentum broadening square of a heavy quark per unit time, $d\langle k_\perp^2\rangle/dt$, due to elastic scatterings and can be derived from Eq.~(\ref{eq:Boltzmann}) as
\begin{align}
\hat{q}=&\sum_{b c d } \frac{\gamma_b}{2E_a}\int  \prod_{i=b,c,d}\frac{d^3p_i}{2E_i(2\pi)^3} f_b |{\cal M}_{ab\rightarrow cd}|^2 S_2(\hat s,\hat t,\hat u)\nonumber\\
 \times & (2\pi)^4\delta^4(p_a\!+\!p_b\!-\!p_c\!-\!p_d) \left[\vec{p}_c-\left(\vec{p}_c\cdot \hat{p}_a\right)\hat{p}_a\right]^2.
\label{eq:qhat}
\end{align}
{This depends on both the temperature through thermal masses and the coupling between quasi-particles $g(T)$ and the heavy quark energy through the coupling between heavy quarks and the gluons $g(E)$.}
Another frequently quoted transport coefficient of heavy quarks is their spatial diffusion coefficient $D_\mathrm{s}$, which can be related to $\hat{q}$ via $D_\mathrm{s}(2\pi T)=8\pi T^3/\hat{q}$. In Fig.~\ref{fig:Ds}, we present this diffusion coefficient as a function of the temperature {for heavy quark energy $E=10$ GeV}. Our constraints are consistent with other model results in the literature~\cite{Song:2015ykw,Scardina:2017ipo,Riek:2010fk,Gossiaux:2008jv,Xu:2017obm} as well as direct lattice QCD calculations~\cite{Ding:2012sp,Banerjee:2011ra,Altenkort:2023oms}. We have also verified that the $\hat{q}$ parameter we obtain is consistent with the constraint provided by the JET Collaboration study~\cite{JET:2013cls} using the inclusive hadron $R_\mathrm{AA}$.

\begin{figure}[tbp]
\includegraphics[width=0.99\linewidth]{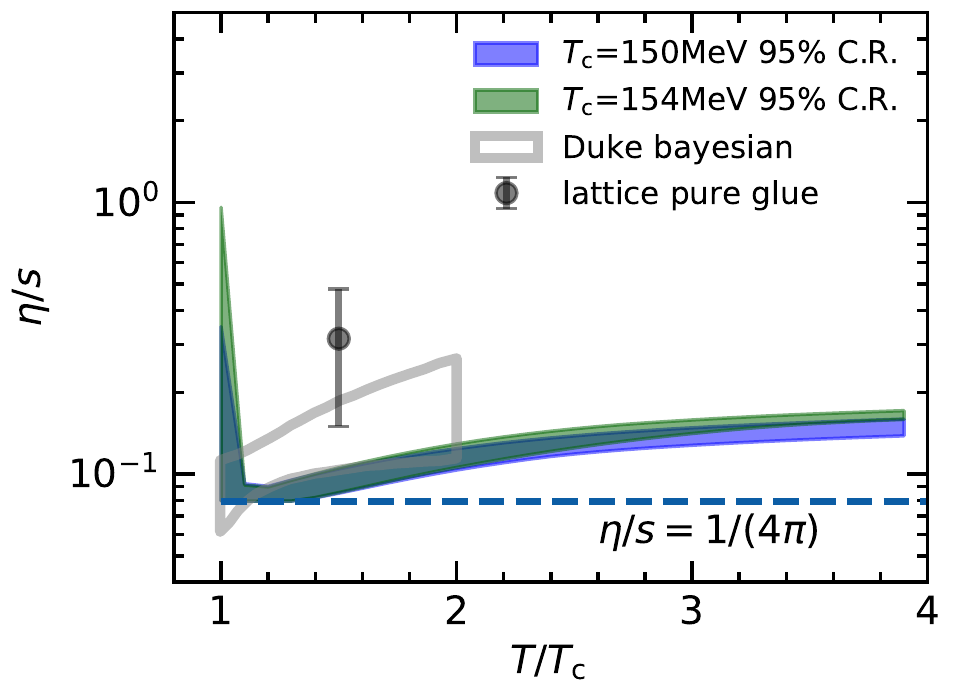}
\caption{(Color online) {The specific shear viscosity ($\eta/s$) as a function of the medium temperature, compared between extractions using two different values of $T_\mathrm{c}$, the Bayesian analysis from the Duke group \cite{Bernhard:2019bmu} and the lattice result for pure glue QCD \cite{Altenkort:2022yhb}.}}
\label{fig:etasratio}
\end{figure}

{With the effective coupling between quasi-particles extracted from the heavy flavor data, one may also calculate the temperature dependent shear viscosity of the quasi-particle system using the relaxation time approximation~\cite{Sasaki:2008fg,Plumari:2011mk}, as shown in Fig. \ref{fig:etasratio}.
{Our result is also compared with the Bayesian analysis from the Duke group \cite{Bernhard:2019bmu} and the lattice result for pure glue QCD \cite{Altenkort:2022yhb}.}
We have verified that little difference is observed in the soft hadron yield and elliptic flow from CLVisc hydrodynamic simulations between using the shear viscosities obtained in the quasi-particle model and a constant value of $\eta/s=0.08$ that was used for hydrodynamic evolution of QGP in this study. In principle, one can also use the quasi-particle model for shear and bulk viscosity in the hydrodynamic simulations and include experimental data on soft hadron spectra in future combined Bayesian analyses of both hard and soft probes. }

{\em Summary.}
Within the QLBT model for jet and heavy quark transport in QGP which evolves according to the CLVisc hydrodynamics and a hybrid fragmentation-coalescence model for hadronization, we have carried out a Bayesian analysis of the experimental data on $D$ meson spectra and anisotropy $v_2$ at both RHIC and LHC. We realized a simultaneous constraint on the properties of the QGP and heavy quark probes. The QGP EOS we extract is consistent with the lattice QCD results, and the heavy quark diffusion coefficient we obtain agrees with results from other models and lattice calculations. In future studies, one can incorporate a more extensive set of parameters, e.g. phase transition temperature $T_\mathrm{c}$, that characterize the medium properties, and involve a broader range of jet observables and soft hadron spectra in order to accomplish the goal of constraining the properties of nuclear matter using hard probes.


{\em Acknowledgments.} We thank W. Ke and L.~G. Pang for helpful discussions. This work is supported in part by NSFC under Grant Nos.~12225503, 11890710, 11890711, 11935007, 12175122, 2021-867, 11221504, 11861131009 and 11890714, by US DOE under Grant No.~DE-AC0205CH11231, and by US NSF under Grant OAC-2004571 within the X-SCAPE Collaboration. Some of the calculations were performed at NSC$^3$.


\bibliography{refs_fll}
\end{document}